\documentclass{rsproca_new}
\jname{rspa}
\Journal{Proc R Soc A\ }

\usepackage{amsmath,amssymb,bm}
\usepackage{url}
\usepackage{isabelle,isabellesym}
\let\ts=\thinspace

\begin{document}


\title*{Computational Logic: Its Origins and Applications}
\author{Lawrence C. Paulson FRS}
\address{Computer Laboratory, University of Cambridge, England, \email{lp15@cam.ac.uk}}

\subject{Theory of computing, mathematical logic}

\keywords{formal verification, theorem proving, proof assistants, Isabelle, LCF}

\corres{Prof.\ Lawrence C. Paulson FRS, University of Cambridge\\
\email{lp15@cam.ac.uk}}

\begin{abstract}
Computational Logic is the use of computers to establish facts in a logical formalism. Originating in 19th-century attempts to understand the nature of mathematical reasoning, the subject now comprises a wide variety of formalisms, techniques and technologies. One strand of work follows the ``LCF approach'' pioneered by Robin Milner FRS, where proofs can be constructed interactively or with the help of users' code (which does not compromise correctness). A refinement of LCF, called Isabelle, retains these advantages while providing flexibility in the choice of logical formalism and much stronger automation. The main application of these techniques has been to prove the correctness of hardware and software systems, but increasingly researchers have been applying them to mathematics itself.
\end{abstract}

\maketitle

\section{Introduction}
\label{sec:intro}

Computers control everything from heart pacemakers to aircraft, putting everyone at risk when they fail. Computer-related failures regularly cause disruption, damage and occasionally death. A recent example is the WannaCry ransomware attack, which hit computers around the world, notably in the National Health Service \cite{woollaston-wannacry}. This attack was made possible by a specific defect, MS17-010, affecting several versions of the Windows operating system. In 1994, a floating-point division error in the Pentium processor forced Intel to recall millions of chips at a cost of \$475 million~\cite{nicely-pentium-fdiv}. Many such failures are caused by carelessness and could be prevented by the introduction of more disciplined development methods. The ultimate discipline is to use formal logic and mathematics to prove the correctness of designs. This is called \textit{formal verification}.

Formal verification typically relates a \textit{formal specification} to a \textit{model implementation}. That is, we start with a complete description of all services to be delivered by the system (the \textit{specification}) and prove that it is satisfied by the model implementation, where both have been expressed in some logical formalism. A proof that all services will be delivered includes a proof that no crashes or other failures will prevent this. However, any formalised implementation must leave out some details of the actual implementation: for example, finite-precision arithmetic may be assumed to be exact, encryption to be unbreakable or hardware components to be infinitely fast. (Readers may remember schoolbook physics problems involving frictionless surfaces, or when friction is included, it is assumed to be linear.) If the model of the implementation is too unrealistic, proofs about it will overlook certain failure modes. No model can include all real-world details, but it is crucial to include the aspects of the implementation that we are most concerned about. For example, in computer security, many systems can be broken without defeating the encryption itself~\cite{lowe-new}. Then it makes sense to assume encryption to be unbreakable if that simplifies the proofs. Validating the model is complementary to the formal proofs and could involve a variety of methods, ranging from professional judgement to systematic testing. (It is a misconception to think that formal proof eliminates the need for testing.)

Formal verification requires the support of specialised computer software. \textit{Computational Logic} is concerned with formal logic as implemented on computers, where it is frequently applied to solve problems related to computation. However, formal verification is increasingly being applied to systems that interact with the real world and deal with phenomena modelled by differential equations, or that have to achieve goals expressed in terms of probability. Therefore, it has become necessary to formalise mathematical knowledge.


\section{A Brief History of Formal Logic}
\label{sec:logic}

A \textit{logical formalism} is a precisely defined symbolic language that includes logical primitives such as ``and'' ($\land$), ``or'' ($\lor$), ``not'' ($\neg$), ``implies'' ($\to$), ``for all'' ($\forall$) and ``there exists'' ($\exists$). It may include the ability to define new symbols as abbreviations for other expressions. It will include rules for making simple logical deductions: for example, from $\phi\land \psi$, the conclusions $\phi$ and $\psi$ can both be derived; from $\phi$ we can derive $\phi\lor \psi$. From such basic primitives, vast swathes of mathematics can be derived.

Formal logic emerged from philosophy in the 19th century. Gottlob Frege's \textit{Begriffschrift} (concept language) was the first recognisably modern treatment of logic. His work introduced the project of reducing the whole of mathematics to logic~\cite{boolos-frege,goedel44}. But Frege's work ended with the discovery of Russell's paradox: 
\begin{quote}
Let $R$ be the set of all sets that are not members of themselves. Then is $R$ a member of itself?	
\end{quote}
If $R$ is a member of itself (written $R\in R$) then it cannot be a member of~$R$, and therefore $R\not\in R$; on the other hand, if $R\not\in R$ then $R\in R$. We get a contradiction either way. Note that as long as we confine ourselves to English, Russell's paradox may resemble amusing puzzles such as ``This statement is false'', but once we use a formal language with strict rules, we are in trouble. For $R\not\in R$ and $R\in R$ are both true, from which all other statements can be proved using basic logic. There are many other paradoxes \cite[p.\ts60]{principia}, e.g.\ Burali-Forti, which involves constructing the greatest ordinal number~$\Omega$, but then $\Omega+1 < \Omega$. As G\"odel remarks,
\begin{quote}
	By analysing the paradoxes to which Cantor's set theory had led, [Russell] freed them from all mathematical technicalities, thus bringing to light the amazing fact that our logical intuitions (i.e.\ intuitions concerning such notions as truth, concept, being, class) are self-contradictory. \cite[p.\ts452]{goedel44}
\end{quote}
From that moment, the development of formal logic was determined by the approach taken to solving the paradoxes. \textit{Set comprehension} --- the idea that every property~$\phi$ created a set, $\{x \mid \phi(x)\}$ --- somehow had to be constrained to deny the existence of $\{x \mid x\not\in x\}$, the Russell set. One approach was to restrict comprehension to some \textit{already existing} set, say~$A$, yielding the smaller set $\{x\in A \mid \phi(x)\}$: the elements of~$A$ satisfying the property~$\phi$. The universe of sets is not itself a set, and it is built up from the empty set using a few primitives such as union and powerset. This approach leads to modern axiomatic set theory \cite{boolos-iterative,kunen80}.

Another solution to the paradoxes involves the notion of~\textit{types}. Whitehead and Russell's original conception of type \cite[p.\ts37]{principia} was rather unclear~\cite{goedel44}, but by 1940, Church \cite{church40} had developed types to a form that survives today. Rather than postulating one universal collection of everything that exists, mathematical objects are classified into distinct types, representing smaller collections. Every construction has a fixed type. When forming a set, all the elements must have the same type, say~$\tau$, and the set itself will have a different type, which we may write as $\tau\,\text{set}$. Thus the fatal proposition $R\in R$ cannot even be written, because if $R$ has type $\tau$ then it cannot have type $\tau\,\text{set}$ at the same time.

Types seem to be unpopular with mathematicians, and it is usual to assert that mathematics is founded on set theory. Whitehead and Russell \cite{principia}, provided no notation or symbols for types, making them invisible, while Church \cite{church40} had one single type~$\iota$ of individuals encompassing all objects that were neither sets nor functions. However, Church's ideas entered into computer science, particularly through his collaboration with Alan Turing, and by the early 1960s, the logical notion of type was starting to get conflated with programming language features (previously called ``modes'') that distinguish between integer and floating-point arithmetic, for example. The ubiquity and utility of types in computer programming may be one reason why computer scientists tend to prefer elaborately typed formalisms.

Yet another solution to the paradoxes was the philosophy of \textit{constructive mathematics} or \textit{intuitionism}~\cite{heyting-foundations}
. Briefly, this is the view that mathematical objects exist only in our imaginations, in contrast to the Platonist or realist view that mathematical objects have their own reality \cite[p.\ts323]{goedel-basic-foundations}. Intuitionism demanded a new treatment of logic, in which the disjunction $\psi\lor\phi$ could be true only if we \textit{knew} which of the two statements, $\psi$ or~$\phi$, was true. Many of the basic tautologies of Boolean logic must then be rejected, starting with $\psi\lor\neg\psi$ (as we have no general way of knowing which alternative holds) and with $\neg\neg\phi$ a weaker statement than~$\phi$. At the level of quantifiers, $\exists x\, \phi(x)$ could be true only if we \textit{knew} some specific value~$a$ for which $\phi(a)$ held.  For that reason, $\exists x\, \phi(x)$ was a stronger claim than $\neg[\forall x\, \neg\phi(x)]$, although in classical logic the two formulas are equivalent.

Intuitionism has not caught on with mainstream mathematicians. However, there are strong links between intuitionistic logic and computation: the knowledge mentioned in the previous paragraph had to be computable. Recent decades have seen the introduction of formalisms~\cite{coquand&huet88,martin-lof-itt-predicative} that identify intuitionistic formulas with elaborate types: every logical symbol has an equivalent at the type level, and to prove a proposition amounts to exhibiting an element of the corresponding type. These formalisms are highly expressive and typically, all expressions denote computable functions. And there is a direct identification between logical propositions and  types \cite{martinlof85}. For example, proving the formula $(\forall x\in A) (\exists y\in B)\, \phi(x,y)$ yields a computable function~$f$ such that if $x\in A$ (that is, $x$ belongs to type~$A$) then $f(x)$ returns some $y\in B$ paired with some sort of ``evidence'' affirming that the property $\phi(x,y)$ holds.%
\footnote{Thus, a form of the axiom of choice is actually provable in this system \cite[p.\ts516]{martinlof85}.}
 Then we would have shown that $f$ had the type $(\prod x\in A) (\sum y\in B)\, \phi(x,y)$, which we can identify with the formula above. Here, $\prod$ and $\sum$ are the type-level counterparts of the quantifiers~$\forall$ and~$\exists$. The elements of $(\prod x\in A)B(x)$ are functions~$f$ that map any given $a\in A$ to some $f(a)\in B(a)$, generalising the function type $A\to B$ by allowing the result type to depend on the value of the argument; similarly, the elements of $(\sum x\in A)B(x)$ are pairs~$\langle a,b\rangle$ where $a\in A$ and $b\in B(a)$, generalising the Cartesian product $A\times B$. This ``propositions as types'' paradigm is the focus of much current research; see Wadler \cite{wadler-propositions} for a detailed historical introduction.

Formal verification today can therefore be done within a wide variety of logical formalisms, which include
\begin{enumerate}
  \item those based on axiomatic set theory, and with no concept of type;
  \item the \textit{simple} type theory of Church, where a few primitive types can be combined using operators such as $\times$ (Cartesian product), $+$ (disjoint sum), $\to$ (function space) and the set type constructor mentioned above;
  \item \textit{dependent} type theories where types can take parameters, as in~$B(x)$, and with variable binding as in $(\prod x\in A)\, B(x)$, thereby allowing the type of $n\times n$ matrices, for example.
\end{enumerate}
The typed approaches have received the most attention in computational logic, and below we shall focus on simple type theory. The third approach is also the focus of much research. 

There are many other formalisms (for example, modal and relevance logics) and computational logic techniques \cite{huth-ryan} not covered below. In particular, \textit{model checking} techniques~\cite{jhala-software-model} are widely used. They can automatically verify system properties expressed in temporal logics (for reasoning about time). However, they rely on state enumeration rather than proof construction, so are out of scope here.

\section{Mechanised Logic: the LCF Tradition}

If we want to prove theorems in a formal logic, computer power is necessary. Whitehead and Russell's development of basic logic, classes and relations only reaches (a precursor to!) 1+1=2 after hundreds of pages \cite[p.\ts360]{principia}. Mathias \cite{Mathias2002} has pointed out that the representation of the number~1 as defined by Bourbaki expands to 4,523,659,424,929 symbols. Better choices can mitigate such issues, but the fact remains that formal proofs are extremely long and detailed.

Ideally, we would like the computer to prove our theorems automatically. If that is impossible, then we would like to give only strategic advice, and let the machine work out the details. As a general rule, greater automation requires the choice of a less expressive formalism. \textit{Boolean logic} (where we only have and, or, not) is decidable, so in principle, all questions can be settled automatically. The problem is known as SAT (for Boolean SATisfiability) and although a general solution is thought to be exponential in the number of Boolean variables, highly efficient heuristic approaches exist. In a dramatic demonstration of this technology, Heule and Kullmann \cite{heule-science-brute} solved the Boolean Pythagorean Triples problem, an open problem in Ramsey Theory, generating a 200TB proof.

Our logic becomes more expressive if we introduce the quantifiers ``for all'' and ``there exists''. If we forbid quantification over functions and sets, this is called \textit{first-order logic}. This theory is \textit{complete}: all theorems can be proved, given unlimited resources. But there is no general way to determine how much effort might be required or whether the claim is false. Automatic theorem proving in first-order logic is a significant field in its own right; one of its main achievements, though 20 years ago, is the solution of the Robbins Conjecture~\cite{mccune-solution}. Incidentally, two early titans of computational logic --- Hilary Putnam \cite{davis-putnam} and J.~Alan Robinson \cite{robinson65} --- were trained as philosophers, and Putnam is chiefly remembered for his essays.

Unfortunately, fully automatic theorem proving cannot meet the needs of verification. Arithmetic, functions and sets are necessary for specifying even simple systems. No automatic procedure can decide arbitrary statements involving integer arithmetic: this is Hilbert's famous \textit{Entscheidungsproblem}, which was settled by Turing~\cite{turing1936a}. Another reason is that specifications may extend to hundreds of pages, and fully automatic theorem proving does not scale well. Interactive methods are the only feasible way to prove theorems based on huge specifications.%
\footnote{Interactive theorem provers are also called \textit{proof assistants}.}
And the name most associated with interactive theorem proving is Robin Milner (FRS 1988)\@.

Milner was interested in an obscure formalism called Logic for Computable Functions (LCF), and built a simple proof checker while at Stanford University. It had two major drawbacks: proofs often required long sequences of repetitive, obvious steps, and the proofs themselves took up too much memory. Upon taking a position at the University of Edinburgh, he created an ambitious new system called Edinburgh LCF \cite{mgordon79}, incorporating several innovations. One was to provide the user with a specialised programming language, called ML (for metalanguage) so that users could extend the system with their own code. That would enable them to deal with any issue of repetitive steps by coding their own automation. Another was to eliminate the storage of proofs through a software technique known as \textit{abstract data types}. By reserving the privilege of creating theorems to a small \textit{kernel} of code, the need to store proofs was eliminated. User-supplied automation, however faulty, could never cause Edinburgh LCF to proclaim an invalid theorem. Edinburgh LCF also introduced an entire framework of techniques and terminology: 
\begin{itemize}
  \item \textit{theory files} containing specifications and definitions of the concepts of interest: theories can build on other theories, allowing hierarchical developments of any size.
  \item \textit{backward proof}: one begins with a statement of the desired result and successively reduces it (using  \textit{proof tactics}) to simpler statements until all have been proved.
  \item \textit{tacticals} for combining proof tactics in simple ways, for example, repetition or executing a series of tactics
  \item built in \textit{automation}, for example to perform routine simplification
  \item along with the ability to \textit{extend} this automation by writing additional ML code.
\end{itemize}
Today, almost all of the main interactive theorem provers incorporate these techniques, and in many cases, actual code from Edinburgh LCF\@. ML turned out to be useful for programming in general, and has had a huge impact on the design of modern programming languages such as Haskell, OCaml and Scala. The meanings of LCF and ML go well beyond their literal meanings as acronyms.%
\footnote{ML can also refer to machine learning.}

Another significant figure in interactive theorem proving was Michael JC Gordon (FRS 1994)\@. He created the HOL system \cite{mgordon-hol}, which survives to this day in several variants, including HOL4 and HOL Light~\cite{hol-light-tutorial}. All of these are descendants of Edinburgh LCF, which he also helped to create~\cite{mgordon79}. Even more significant is Gordon's development of techniques for verifying computer hardware; when he started working on this topic in the early 1980s, essentially all verification research was focused on software. Gordon developed techniques that scaled from a single-transistor device to an entire computer. His key insight was that the right formalism for this task was \textit{simple type theory}, also known as \textit{higher-order logic}, hence HOL\@. 

The choice of higher-order logic was radical at the time. First-order logic was dominant, thanks perhaps to its strong theoretical properties such as completeness. Gordon noted that such properties were irrelevant to verification, and recognised that the additional expressiveness of higher-order logic was necessary~\cite{mgordon86}. His choice represented a return to the stronger logics of the early twentieth century. With his students, Gordon extended his techniques to cover a variety of digital systems, including floating-point hardware, probabilistic algorithms, and many other applications. Typical of these is Harrison's verification of a floating-point algorithm for the computation of the exponential function~\cite{harrison-exp}. The proof required a complete formalisation of the IEEE floating-point standard, in all its complexity, as well as the error analysis for the algorithm itself.

While on the topic of radical choices, we must not overlook research into constructive type theories. Inspired originally by the work of Martin-L\"of \cite{martinlof85} and by Coquand and Huet~\cite{coquand&huet88}, researchers built a series of interactive theorem provers, all following the Edinburgh LCF paradigm. For reasons of space, I will only mention Coq~\cite{coq-book}, which has become perhaps the most popular interactive theorem prover in the world. Typical of all these approaches is that through dependent types, the type system carries out much of the reasoning burden that would otherwise be done by explicit proof steps. Among the landmark achievements carried out using Coq are the formal verification of a C compiler \cite{leroy-compiler} and the formalisation of the odd order theorem~\cite{gonthier-oot}. A full treatment of this line of work would require a separate article.
 
\section{A New Theorem-Proving Architecture: Isabelle}

During the 1980s, many different logical formalisms were competing for researchers' attention. In addition to LCF,  type theories were evolving rapidly, along with a wide variety of formalisms for proving software correct. One drawback of the LCF approach was its requirement that every primitive inference rule had to be implemented as program code, and then a second time as a proof tactic. An error in the latter code could prevent the user from obtaining the theorem they thought they had proved, while an error in the former code could allow the system to assert false statements as true. 

To understand this drawback in more detail, consider one of the simplest logical inference rules: if $\phi$ and~$\psi$ are theorems, then so is their conjunction, $\phi\land\psi$. It is often written as
\[ \frac{\phi \quad \psi}{\phi\land\psi.} \]
However, note that mathematical proofs frequently involve temporary assumptions, such as when we consider the two cases of whether some integer~$n$ is even or odd. Each of the two cases is proved with the help of the corresponding assumption about~$n$.
So people generally adopt a so-called \textit{natural deduction} calculus, where all reasoning takes place within a \textit{context}: a set of formulas $\theta_1$, \ldots, $\theta_n$ temporarily assumed to be true.%
\footnote{Mathematical proofs also introduce temporary quantities, as when we allow~$x$ to denote a root of some polynomial. Therefore contexts may include a string of bound variables, but the details~\cite{paulson-found} are too complicated to present here.}
In order to minimise the need for subscripting, we typically use the large Greek letters $\Gamma$ and~$\Delta$ to stand for contexts. If we chose to make this explicit, our inference rule could be written (forming what is known as a \textit{sequent calculus})
\[ \frac{\Gamma \Longrightarrow \phi \qquad \Delta \Longrightarrow \psi}{\Gamma,\Delta \Longrightarrow \phi\land\psi.} \]
Note that the two contexts are combined when we form the conclusion, since it depends on everything assumed by the premises.  And this is the simplest possible example. A number of rules modify contexts by ``discharging'' assumptions. The treatment of quantified variables introduces further technicalities.

To implement this rule using the LCF architecture (which is also adopted by the HOL family and Coq), we begin by designing data structures that can represent the full syntax of our formalism. Then we collect all the inference rules that need to be implemented; taken together, they will define the abstract data type \texttt{thm}, the type of theorems. Each rule will take the form of a function coded in ML\@. The function for the rule above will take as its arguments the two premises, namely $\Gamma \Longrightarrow \phi$ and $\Delta \Longrightarrow \psi$. This rule is simple, but in other cases, the code must make syntactic checks  on the premises to ensure that they have the correct form, and if not, signal an error. Finally, the function returns $\Gamma,\Delta \Longrightarrow \phi\land\psi$ as a value of type \texttt{thm}, thus certifying it to be a theorem. This particular function is called \texttt{CONJ} in LCF and its descendants\@.

Defining type \texttt{thm} as outlined above implements a logic in ML, but a usable system must also support backward proof: the ability to prove a statement of the form $\phi\land\psi$ by proving the two parts separately. So we need to write another function, called a \textit{tactic}, which essentially executes the inference rule in reverse. Given the goal of the form $\Gamma \Longrightarrow \phi\land\psi$, it returns the two separate subgoals $\Gamma \Longrightarrow \phi$ and $\Gamma \Longrightarrow \psi$ (and note that here we have one single context, $\Gamma$, since we are working backwards). 

These subgoals are not theorems but merely statements to work on next. In order to yield a theorem at the end, the tactic also needs to return a piece of code involving \texttt{CONJ}\@. This theorem-yielding function is called the \textit{validation}, and the design allows most of the code to lie outside the abstract data type \texttt{thm}, on which the soundness of the whole system depends. If the tactic or its validation is somehow wrong, the system's soundness will not be compromised. However, the user is likely to be disappointed, making the effort to prove the theorem but never getting the theorem itself. (Instead of an error will be signalled, or possibly some other theorem delivered.) Gordon's retrospective on these techniques is informative~\cite{gordon-tactics-milner}, as is an early exposition of mine~\cite{paulson-natural}.

So we see that to implement any formal calculus using the LCF architecture, every inference rule must be implemented in code twice. There could be dozens of rules, some of them complicated, and any error would be damaging. Modifying an LCF-style system to implement a new formalism required months of work.

The solution to these problems is to use the LCF idea in a different way. Instead of focusing on the theorems of the given formalism, we should focus on \textit{contextual implications} of the form $\Gamma \Longrightarrow \phi$. In particular, such implications can be joined end-to-end, creating proof trees. This approach has a number of advantages: 
\begin{itemize}
  \item Contexts are managed in the same way for every formalism being supported. Meanwhile, formalisms that manage the context explicitly can still be handled.
  \item The incremental buildup of a proof tree can express both forward and backward proof.
  \item Subgoals can even contain shared ``placeholder'' variables, as may arise when proving something like $\exists x\, (\phi(x)\land\psi(x))$. That is to say, we can try to prove something of the form $\phi(-)$ and if we manage to prove $\phi(f(1))$, then the other subgoal will become $\psi(f(1))$.
\end{itemize}
But the main advantage is simply that inference rules can now be written as assertions.  For example, the conjunction rule above becomes
\[ \phi, \psi \Longrightarrow \phi\land\psi. \]
The rule to eliminate a disjunction (proof by case analysis) becomes
\[ \phi\lor\psi, (\phi \Longrightarrow\theta), (\psi\Longrightarrow\theta) \Longrightarrow \theta. \]
The rule to introduce an existential quantifier becomes
\[ \phi(a) \Longrightarrow \exists x\,\phi(x). \]
This architecture allows primitive inferences of a formalism to be entered more or less as they might appear in a textbook. 

Isabelle is the world's first \textit{generic} theorem prover: designed to support logical inference in general as opposed to any single formalism. Isabelle takes some ideas from higher-order logic to represent formal syntax and logical inference rules~\cite{paulson700}.  Notational features such as variable binding (the $i$ in $\sum_{i\in I}\,t_i$) and principles such as mathematical induction could be expressed straightforwardly, and conformance to traditional presentations could be proved~\cite{paulson-found}. Such meta-logical systems are known as \textit{logical frameworks}~\cite{harper-jacm}, the first of which was the AUTOMATH language created by N. G. de~Bruijn \cite{debruijn-survey}.

Another fundamental Isabelle principle is to provide the basic components of automatic proof search at the lowest level. The resolution techniques of first-order logic had already shown promise. It seemed right to adopt some of those techniques to provide automation for other formalisms. The joining together of implications, which is Isabelle's core inference mechanism, is itself a form of resolution. Over the years, increasingly sophisticated forms of automatic proof search were added to Isabelle~\cite{paulson-blast}. 

Have these ideas stood the test of time? Certainly Isabelle is the only widely-used interactive theorem prover to be based on a logical framework. Meanwhile, efforts to exploit the power of resolution theorem proving finally resulted in strong integration of the two technologies: the \textit{sledgehammer} tool~\cite{paulson-three-years} applies a battery of resolution theorem provers to a problem. It often finds a proof with the help of obscure facts from Isabelle's extensive libraries. Some of the most impressive work done using Isabelle---from the verification of security protocols~\cite{paulson-tls} to a verified operating system kernel~\cite{klein-sel4} to formalised mathematical theorems \cite{avigad-clt,paulson-incompl-logic}---was only possible thanks to Isabelle's automation.

On the other hand, the idea of generic theorem proving is less successful. It's true that Isabelle has been used to support a great many formalisms~\cite{paulson700}, including a version of  Martin-L\"of's intuitionistic type theory, first-order logic, set theory and some modal logics. Isabelle/ZF develops a substantial amount of Zermelo-Fraenkel set theory~\cite{paulson-consistency}. One can argue that, even in the case of a single formalism, the use of a logical framework makes for a more flexible notion of proof construction than that found in traditional LCF-like systems. However, Isabelle's facility to support multiple formalisms is hardly used. Its most popular instantiation by far is Isabelle/HOL, which at bottom is yet another implementation of Church's higher-order logic \cite{church40}. Constructive type theories tend to be demanding syntactically; that research community has always preferred to build its own tools.

Isabelle's core ideas were in place by 1989~\cite{paulson-found}. Since then, Isabelle development has increasingly moved to the Technical University of Munich, under the leadership of Prof.\ Tobias Nipkow.  Of the innumerable contributions made there, the most important one is the Isar structured proof language~\cite{wenzel-isabelle/isar}, which although completely formal includes many mathematical idioms and often allows legible, natural-looking proofs. (Hitherto, Isabelle followed LCF in expecting users to generate proofs directly through the ML programming language.) Extensions to Isabelle's core ideas to support higher-order logic are also due to Nipkow and his colleagues, in particular, the notion of type classes~\cite{wenzel-type}. The years have seen a long series of refinements made and case studies undertaken. One notable example is counterexample detection~\cite{blanchette-nitpick}, whereby Isabelle can frequently inform the user that the statement under consideration is not a theorem.

\section{Formalising Mathematics}

The idea of performing mathematics by machine goes back at least as far as Leibniz, who invented a calculating machine. The first mathematician to properly realise this dream was de~Bruijn \cite{debruijn-survey}, whose AUTOMATH project started in the 1960s and culminated in the formalisation of a monograph on the construction of the reals~\cite{jutting77}. De Bruijn's approach was based on dependent type theories of his own devising. At around the same time, Andrzej Trybulec had the idea of formalising mathematics using a form of typed set theory; a substantial amount of mathematics was translated into his Mizar formalism, with a particularly elegant treatment of algebraic structures~\cite{rudnicki-algebra}.

However, the main impetus for formalised mathematics has arisen from the requirements of verification, which has been a priority area for research funding. Harrison's early work~\cite{harrison-exp} to verify a floating-point algorithm for the exponential function naturally led to the formalisation of increasing amounts of analysis. In order to verify probabilistic algorithms (which are controlled by virtual coin tosses), Hurd~\cite{hurd-primality} formalised some probability theory, including  measure theory and Lebesgue integration. Recent work in this direction includes H\"olzl's comprehensive formalisation~\cite{holzl-markov} of Markov chains, an abstract model of probabilistic systems that has numerous applications. Such work has generally been done using versions of HOL and Isabelle/HOL, but floating-point algorithms have also been verified using Coq~\cite{boldo-fpexpansion}.

Over time, competition arose between the proponents of different verification systems on the basis of how many well-known theorems had been formalised in each. Harrison took the lead here, formalising numerous results in his HOL Light system~\cite{hol-light-tutorial}. These included much of complex analysis~\cite{harrison-complex} as well as a proof of the prime number theorem~\cite{harrison-pnt} and many other results. Later, Harrison played a major role in the effort to formally verify Hales's 1998 proof \cite{hales-kepler} of the Kepler Conjecture, a 400-year-old problem concerning the optimal packing of spheres. Referees had objected to this proof because of its reliance on extensive computer calculations. In response, Hales launched the Flyspeck project~\cite{hales-formal-Kepler} to formally verify his proof. Flyspeck was ultimately successful, confirming and simplifying Hales's argument~\cite{hales-revision-proof}. Some of the formal proofs were done using Isabelle~\cite{nipkow-flyspeck-tame}.

This was not the first time a mathematical result was formalised in order to confirm a contested proof. In 2005, Gonthier~\cite{gonthier-4ct} formalised the Four Colour Theorem within Coq. The 1976 proof by Appel and Haken was notorious for its the use of a bespoke computer program to check nearly 2000 cases. Gonthier used a similar strategy, with a crucial difference: the cases were checked by algorithms that had been verified using Coq.

Other work of this sort concerned proofs that, while enjoying the full confidence of mathematicians, had other issues. My own formalisation of G\"odel's second incompleteness theorem~\cite{paulson-incompl-logic} falls into this category: most published proofs are sorely lacking in detail. Fleuriot's formalisation~\cite{fleuriot-book} of some material from Newton's \textit{Principia} is remarkable for embracing Newton's use of infinitesimals: Fleuriot formally justifies the original proofs through \textit{non-standard analysis}, which gives a rigorous foundation to infinitesimal reasoning. Above all, we have Gonthier's monumental project to verify the odd order theorem~\cite{gonthier-oot}; the issue here is simply size, as the theorem was originally proved in a 255-page journal article. 

Generally speaking, the point of formalised mathematics is to clear up doubts, resolve ambiguities and identify errors. Mathematicians regularly make mistakes, as examined by Lakatos~\cite{lakatos} in his history of Euler's polyhedron formula; he points out that even definitions can be wrong. Famous modern mathematicians can also be fallible:
\begin{quote}
  When the Germans were planning to publish Hilbert's collected papers and to present him with a set on the occasion of one of his later birthdays, they realised that they could not publish the papers in their original versions because they were full of errors, some of them quite serious. Thereupon they hired a young unemployed mathematician, Olga Taussky-Todd, to go over Hilbert's papers and correct all mistakes. Olga laboured for three years. \cite[p.\ts201]{rota-indiscrete-thoughts}. 
\end{quote}
Increasingly, the impetus for formalising mathematics comes from mathematicians themselves. Fields medallist Vladimir Voevodsky wrote of his horror at finding errors in his own work: 
\begin{quote}
	And who would ensure that I did not forget something and did not make a mistake, if even the mistakes in much more simple arguments take years to uncover?  \cite{voevodsky-origins}
\end{quote}
Voevodsky's \textit{homotopy type theory}~\cite{hottbook} is a new approach to the formalisation of mathematics. It gives Martin-L\"of's intuitionistic type theory a new interpretation involving topological spaces, where types denote homotopy types.%
\footnote{A \textit{homotopy} between two continuous functions $f$ and $g$ is a continuous ``deformation'' of $f$ into~$g$.
With this concept one can define an equivalence relation on topological spaces, and the resulting equivalence classes are called \textit{homotopy types}.}
Moreover, this interpretation suggests the new \textit{univalence axiom}, which says that isomorphic constructions can be identified~\cite{awodey-univalence}. These ideas have generated widespread excitement, and experiments are under way using the Coq proof assistant. 

In the opposite direction, axiomatic set theory has also been used to formalise mathematics. Sets and types both denote collections, with one great difference: typically a variable can have only one type but can belong to any number of sets. So set theory may offer the greatest flexibility in formalising mathematics. Work of my own concerns proving a number of properties to be equivalent to the axiom of choice~\cite{paulson-gr} as well as formalising G\"odel's proof of the relative consistency of the axiom of choice~\cite{paulson-consistency}.  This work can be criticised as being only about set theory itself. However, recent work by Zhan shows that set theory can also be used to formalise traditional mathematical results from group theory, point-set topology and real analysis~\cite{zhan-fundamental}.

\section{Obstacles to Formalising Mathematics}

The obstacles to the formalisation of mathematics are perhaps not what people imagine. G\"odel's theorem is not especially relevant to mathematics in practice. Even the \textit{Entscheidungsproblem} is not that relevant: when we do have decision procedures, they can be too expensive to actually run. Anyway, our ambition is to \textit{assist} mathematicians, not to replace them by computers. The obstacles we face take a variety of forms.

\textit{Foundational}. A mathematician's imagination is unconstrained, but every formal calculus is a rigid framework and ultimately a prison. For example, axiomatic set theory begins by creating a world big enough for 19th-century mathematics, then escalates it through powerful principles to create an inconceivably large universe of mathematical objects. It contains such oddities as the limit of $\{\aleph_0, \aleph_{\aleph_0}, \aleph_{\aleph_{\aleph_0}}, \ldots\}$, a cardinal $\lambda$ satisfying the equation $\lambda = \aleph_\lambda$. (The $\aleph$-sequence enumerates the infinite cardinalities, starting with $\aleph_0$: the cardinality of the integers.) Boolos \cite{boolos-must-believe} frankly states that he can't believe that such an object exists. And yet, category theory \cite{maclane} begins by creating the category of sets and functions, assuming at a stroke the existence of a mathematical object incorporating the entire universe of sets, and building upon it. So one moment our universe seems far too large, and a moment later, it has become too small.

\textit{Conceptual}. Intuition plays a huge role in mathematics. A mathematician easily sees that a certain function --- perhaps defined by a complicated integral --- is continuous. Many convergence properties are obvious. But in some cases, the formal proofs are enormous~\cite{harrison-pnt}. There are a variety of responses to this situation. One is to provide additional automation to deal with specific tasks, such as proving asymptotic properties~\cite{eberl-asymptotics}. Another is to accept that some apparently obvious facts, such as the Jordan curve theorem,%
\footnote{which states that every simple closed curve partitions the plane into an interior and an exterior region}
really are extremely difficult to prove rigorously~\cite{hales-jordan-curve}. Another is to accept that many apparently obvious statements are actually false~\cite{lakatos}. But for all that, the use of any logical formalism requires immense patience.

\textit{Notational}. The language of mathematics has evolved over centuries. It can be inconsistent and ambiguous --- compare $\sin^2 x$ with $y^2x$ --- but it is familiar and reasonably effective. Any formalism implemented on a computer will be code, and look like it. For example, the fundamental theorem of algebra states that if $n$ is a positive integer and $a_0$, $a_1$, \ldots, $a_n$ are complex numbers with $a_n\not=0$, then $a_nz^n +\cdots+a_1z+a_0 = 0$ for some complex~$z$. The HOL Light version relaxes the precondition to $a_0=0$ or otherwise that the~$a_k$ are not all zero for $k=1,\ldots,n$:
\begin{verbatim}
!a n. a(0) = Cx(&0) \/ ~(!k. k IN 1..n ==> a(k) = Cx(&0))
         ==> ?z. vsum(0..n) (\i. a(i) * z pow i) = Cx(&0)
\end{verbatim}
The elegance of the mathematical notation is completely lost. Isabelle does a bit better, allowing the Greek alphabet and many mathematical symbols, but all such languages are still code.

One further complication deserves a mention. Many expressions of mathematics can be \textit{undefined}, e.g.\ some limits, derivatives, integrals or simply $x/y$ when $y=0$. A formal calculus can deal with this problem in a number of different ways~\cite[\S5.1]{avigad-clt}. So we may find that $x/y=x/y$ (when $x/0$ denotes some arbitrary value) or that $x/y=x/y$ if and only if $y\not=0$ (when we regard primitive assertions about undefined values as necessarily false) or we may find that the very act of writing $x/y$ is forbidden unless we can prove $y\not=0$. This last approach can be inconvenient \cite[\S4]{boldo-coquelicot}, indeed each approach has its good and bad points. The idea that $x/y$ is always something horrifies some mathematicians (especially if we go further, as in Isabelle, and actually stipulate that $x/0=0$).  Elaborate treatments of definedness yield greater expressive power: paid for, as usual, by more complicated notations and harder proofs.

\section{The Way Forward}

In the course of this paper we have narrowed our focus several times, deliberately seeking out research as opposed to applications. This is necessary because computational logic tools, interpreted broadly, are now widely used in the development of hardware, software and in the provision of online services. We are in danger, perhaps, of overlooking the main motivation for computational logic: proving the correctness of computer systems. So let's take a moment to consider a basic problem in computer science: compiler correctness.

A compiler is a piece of software that translates program code (written in a programming language such as C or Java) into some sort of binary form that can be executed on computer hardware. Compilers are complex, not least because they use a wide variety of sophisticated techniques to generate binary code that will run as fast as possible. If the compiler is somehow faulty then it could introduce errors into the code that is actually executed. People have sought to prove compilers correct using computational logic techniques~\cite{leroy-compiler}. But even if we finally obtain a correct compiler, how do we translate it into executable binary code? A further complication is that the semantics of a language like C has little in common with the way C is used in real software, which relies on innumerable undefined behaviours.

Kumar et al.~\cite{kumar-cakeml-verified} have solved this chicken-and-egg problem. They have designed their own programming language, CakeML, with a clean semantics. They express a simple compiling algorithm as a mathematical function in HOL4 (a descendant of Gordon's original HOL system). They translate the same algorithm to CakeML and compile it to binary (for the popular x86 architecture) \textit{using HOL4 itself}, giving high assurance of correctness. This yields a trustworthy version of the compiler in binary form. Their approach is an instance of the well-known technique of \textit{bootstrapping}, where a compiler is built for a simple but adequate language and then used as a tool to further its own development until the full language is supported. In this case, the bootstrapping process includes verification, and is possible because HOL4 can cope with the full binary code of the CakeML compiler. This is one representative of a large body of research concerned with treating mathematical functions defined within a theorem prover as executable code.

Readers who want to pursue these topics further will find surveys available on formally verified mathematics \cite{avigad-formally-verified} and on the history of interactive theorem proving \cite{geuvers-proof-assistants,harrison-history}. Isabelle can be downloaded via the URL \url{https://isabelle.in.tum.de}.

\paragraph*{Data accessibility.}
No experimental data involved. There is a software download link above.

\paragraph*{Competing interests.}
Not applicable.

\paragraph*{Funding.}
Much of the research reported here was supported by the EPSRC or its predecessors, or by German and EU funding agencies, going back 40 years. 

\paragraph*{Acknowledgements.}
Angeliki Koutsoukou-Argyraki commented on a draft of this paper. The referees also made insightful comments. I would also like to thank Mike Gordon FRS and Robin Milner FRS for their mentorship.


\begin{thebibliography}{10}

\bibitem{avigad-formally-verified}
J.~Avigad and J.~Harrison.
\newblock Formally verified mathematics.
\newblock {\em Commun. ACM}, 57(4):66--75, Apr. 2014.

\bibitem{avigad-clt}
J.~Avigad, J.~H{\"o}lzl, and L.~Serafin.
\newblock A formally verified proof of the central limit theorem.
\newblock {\em Journal of Automated Reasoning}, 59(4):389--423, Dec. 2017.

\bibitem{awodey-univalence}
S.~Awodey, {\'A}.~Pelayo, and M.~A. Warren.
\newblock Voevodsky's univalence axiom in homotopy type theory.
\newblock {\em Notices of the AMS}, 60(9):1164--1167, Oct. 2013.

\bibitem{itp-2017}
M.~Ayala-Rinc{\'o}n and C.~A. Mu{\~{n}}oz, editors.
\newblock {\em Interactive Theorem Proving ---- 8th International Conference,
  ITP 2017, Bras{\'\i}lia, Brazil, September 26--29, 2017, Proceedings}.
  Springer International Publishing, 2017.

\bibitem{ben-putnam}
P.~Benacerraf and H.~Putnam, editors.
\newblock {\em Philosophy of Mathematics: Selected Readings}.
\newblock Cambridge University Press, 2nd edition, 1983.

\bibitem{coq-book}
Y.~Bertot and P.~{Cast\'eran}.
\newblock {\em Interactive Theorem Proving and Program Development: Coq'Art:
  The Calculus of Inductive Constructions}.
\newblock Springer, 2004.

\bibitem{blanchette-nitpick}
J.~C. Blanchette and T.~Nipkow.
\newblock {Nitpick}: A counterexample generator for higher-order logic based on
  a relational model finder.
\newblock In M.~Kaufmann and L.~C. Paulson, editors, {\em Interactive Theorem
  Proving}, volume 6172 of {\em Lecture Notes in Computer Science}, pages
  131--146. Springer, 2010.

\bibitem{boldo-fpexpansion}
S.~Boldo, M.~Joldes, J.~Muller, and V.~Popescu.
\newblock Formal verification of a floating-point expansion renormalization
  algorithm.
\newblock In Ayala-Rinc{\'o}n and Mu{\~{n}}oz \cite{itp-2017}, pages 98--113.

\bibitem{boldo-coquelicot}
S.~Boldo, C.~Lelay, and G.~Melquiond.
\newblock Coquelicot: {A} user-friendly library of real analysis for {Coq}.
\newblock {\em Mathematics in Computer Science}, 9(1):41--62, 2015.

\bibitem{boolos-frege}
G.~S. Boolos.
\newblock {Gottlob Frege} and the foundations of arithmetic.
\newblock In {\em Logic, Logic, and Logic\/} \cite{boolos-logic}, pages
  143--154.

\bibitem{boolos-iterative}
G.~S. Boolos.
\newblock The iterative conception of set.
\newblock In {\em Logic, Logic, and Logic\/} \cite{boolos-logic}, pages 13--29.

\bibitem{boolos-logic}
G.~S. Boolos.
\newblock {\em Logic, Logic, and Logic}.
\newblock Harvard University Press, 1998.

\bibitem{boolos-must-believe}
G.~S. Boolos.
\newblock Must we believe in set theory?
\newblock In {\em Logic, Logic, and Logic\/} \cite{boolos-logic}, pages
  120--132.

\bibitem{church40}
A.~Church.
\newblock A formulation of the simple theory of types.
\newblock {\em Journal of Symbolic Logic}, 5:56--68, 1940.

\bibitem{coquand&huet88}
T.~Coquand and G.~Huet.
\newblock The calculus of constructions.
\newblock {\em Information and Computation}, 76:95--120, 1988.

\bibitem{davis-putnam}
M.~Davis and H.~Putnam.
\newblock A computing procedure for quantification theory.
\newblock {\em J.~ACM}, 7(3):207--215, July 1960.

\bibitem{debruijn-survey}
N.~G. de~Bruijn.
\newblock A survey of the project {AUTOMATH}.
\newblock In J.~Seldin and J.~Hindley, editors, {\em To H.B. Curry: Essays in
  Combinatory Logic, Lambda Calculus and Formalism}, pages 579--606. Academic
  Press, 1980.

\bibitem{eberl-asymptotics}
M.~Eberl.
\newblock Semi-automatic asymptotics in {Isabelle/HOL}.
\newblock Online at \url{https://www.newton.ac.uk/seminar/20170705133014302},
  July 2017.
\newblock Presented at the Isaac Newton Institute for Mathematical Sciences.

\bibitem{fleuriot-book}
J.~D. Fleuriot.
\newblock {\em A Combination of Geometry Theorem Proving and Nonstandard
  Analysis, with Application to {Newton's} {Principia}}.
\newblock Springer, 2001.

\bibitem{geuvers-proof-assistants}
H.~Geuvers.
\newblock Proof assistants: History, ideas and future.
\newblock {\em Sadhana}, 34(1):3--25, Feb 2009.

\bibitem{goedel44}
K.~G{\"o}del.
\newblock Russell's mathematical logic.
\newblock In Benacerraf and Putnam \cite{ben-putnam}, pages 447--469.
\newblock First published in 1944.

\bibitem{goedel-basic-foundations}
K.~G{\"o}del.
\newblock Some basic theorems on the foundations of mathematics and their
  implications.
\newblock In S.~Feferman, editor, {\em {Kurt G\"odel}: Collected Works}, volume
  III, pages 304--323. Oxford University Press, 1995.

\bibitem{gonthier-4ct}
G.~Gonthier.
\newblock The four colour theorem: Engineering of a formal proof.
\newblock In D.~Kapur, editor, {\em Computer Mathematics}, LNCS 5081, pages
  333--333. Springer Berlin Heidelberg, 2008.

\bibitem{gonthier-oot}
G.~Gonthier, A.~Asperti, J.~Avigad, Y.~Bertot, C.~Cohen, F.~Garillot,
  S.~Le~Roux, A.~Mahboubi, R.~O'Connor, S.~Ould~Biha, I.~Pasca, L.~Rideau,
  A.~Solovyev, E.~Tassi, and L.~Th{\'e}ry.
\newblock A machine-checked proof of the odd order theorem.
\newblock In S.~Blazy, C.~Paulin-Mohring, and D.~Pichardie, editors, {\em
  Interactive Theorem Proving}, LNCS 7998, pages 163--179. Springer, 2013.

\bibitem{mgordon86}
M.~J.~C. Gordon.
\newblock Why higher-order logic is a good formalism for specifying and
  verifying hardware.
\newblock In G.~Milne and P.~A. Subrahmanyam, editors, {\em Formal Aspects of
  {VLSI} Design}, pages 153--177. North-Holland, 1986.

\bibitem{gordon-tactics-milner}
M.~J.~C. Gordon.
\newblock Tactics for mechanized reasoning: A commentary on {Milner} (1984)
  {\textquoteleft}the use of machines to assist in rigorous
  proof{\textquoteright}.
\newblock {\em Philosophical Transactions of the Royal Society of London A:
  Mathematical, Physical and Engineering Sciences}, 373(2039), 2015.

\bibitem{mgordon-hol}
M.~J.~C. Gordon and T.~F. Melham.
\newblock {\em Introduction to {HOL}: A Theorem Proving Environment for Higher
  Order Logic}.
\newblock Cambridge University Press, 1993.

\bibitem{mgordon79}
M.~J.~C. Gordon, R.~Milner, and C.~P. Wadsworth.
\newblock {\em Edinburgh {LCF}: A Mechanised Logic of Computation}.
\newblock LNCS 78. Springer, 1979.

\bibitem{hales-kepler}
T.~C. {Hales}.
\newblock {An overview of the Kepler conjecture}.
\newblock {\em ArXiv Mathematics e-prints}, Nov. 1998.

\bibitem{hales-jordan-curve}
T.~C. Hales.
\newblock The {Jordan} curve theorem, formally and informally.
\newblock {\em The American Mathematical Monthly}, 114(10):882--894, Dec. 2007.

\bibitem{hales-formal-Kepler}
T.~C. Hales et~al.
\newblock A formal proof of the {Kepler} conjecture.
\newblock {\em arXiv.org}, abs/1501.02155, Jan. 2015.

\bibitem{hales-revision-proof}
T.~C. Hales, J.~Harrison, S.~McLaughlin, T.~Nipkow, S.~Obua, and R.~Zumkeller.
\newblock A revision of the proof of the {Kepler} conjecture.
\newblock {\em Discrete {\&} Computational Geometry}, 44(1):1--34, 2010.

\bibitem{harper-jacm}
R.~Harper, F.~Honsell, and G.~Plotkin.
\newblock A framework for defining logics.
\newblock {\em J.~ACM}, 40(1):143--184, 1993.

\bibitem{hol-light-tutorial}
J.~Harrison.
\newblock {HOL Light}: A tutorial introduction.
\newblock In M.~K. Srivas and A.~J. Camilleri, editors, {\em Formal Methods in
  Computer-Aided Design: FMCAD '96}, LNCS 1166, pages 265--269. Springer, 1996.

\bibitem{harrison-exp}
J.~Harrison.
\newblock Floating point verification in {HOL} {L}ight: the exponential
  function.
\newblock {\em Formal Methods in System Design}, 16:271--305, 2000.

\bibitem{harrison-complex}
J.~Harrison.
\newblock Formalizing basic complex analysis.
\newblock In R.~Matuszewski and A.~Zalewska, editors, {\em From Insight to
  Proof: Festschrift in Honour of Andrzej Trybulec}, volume 10(23) of {\em
  Studies in Logic, Grammar and Rhetoric}, pages 151--165. University of
  Bia{\l}ystok, 2007.

\bibitem{harrison-pnt}
J.~Harrison.
\newblock Formalizing an analytic proof of the prime number theorem.
\newblock {\em Journal of Automated Reasoning}, 43(3):243--261, 2009.

\bibitem{harrison-history}
J.~Harrison, J.~Urban, and F.~Wiedijk.
\newblock History of interactive theorem proving.
\newblock In J.~Siekmann, editor, {\em Handbook of the History of Logic
  (Computational Logic)}, volume~9, pages 135--214. Elsevier, 2014.

\bibitem{heule-science-brute}
M.~J.~H. Heule and O.~Kullmann.
\newblock The science of brute force.
\newblock {\em Commun. ACM}, 60(8):70--79, July 2017.

\bibitem{heyting-foundations}
A.~Heyting.
\newblock The intuitionist foundations of mathematics.
\newblock In Benacerraf and Putnam \cite{ben-putnam}, pages 52--61.
\newblock First published in 1944.

\bibitem{holzl-markov}
J.~H{\"o}lzl.
\newblock {Markov} chains and {Markov} decision processes in {Isabelle/HOL}.
\newblock {\em Journal of Automated Reasoning}, 59(3):345--387, Oct. 2017.

\bibitem{hurd-primality}
J.~Hurd.
\newblock Verification of the {Miller-Rabin} probabilistic primality test.
\newblock {\em Journal of Logic and Algebraic Programming}, 56:3--21, 2002.

\bibitem{huth-ryan}
M.~Huth and M.~Ryan.
\newblock {\em Logic in Computer Science: Modelling and Reasoning about
  Systems}.
\newblock Cambridge University Press, 2nd edition, 2004.

\bibitem{jhala-software-model}
R.~Jhala and R.~Majumdar.
\newblock Software model checking.
\newblock {\em ACM Comput. Surv.}, 41(4):21:1--21:54, Oct. 2009.

\bibitem{jutting77}
L.~Jutting.
\newblock {\em Checking {Landau's} ``{Grundlagen}'' in the {AUTOMATH} System}.
\newblock PhD thesis, Eindhoven University of Technology, 1977.

\bibitem{klein-sel4}
G.~Klein, J.~Andronick, K.~Elphinstone, G.~Heiser, D.~Cock, P.~Derrin,
  D.~Elkaduwe, K.~Engelhardt, R.~Kolanski, M.~Norrish, T.~Sewell, H.~Tuch, and
  S.~Winwood.
\newblock sel4: Formal verification of an operating-system kernel.
\newblock {\em Commun. ACM}, 53(6):107--115, June 2010.

\bibitem{kumar-cakeml-verified}
R.~Kumar, M.~O. Myreen, M.~Norrish, and S.~Owens.
\newblock {CakeML}: A verified implementation of {ML}.
\newblock In {\em ACM SIGPLAN-SIGACT Symposium on Principles of Programming
  Languages}, POPL '14, pages 179--191. ACM, 2014.

\bibitem{kunen80}
K.~Kunen.
\newblock {\em Set Theory: An Introduction to Independence Proofs}.
\newblock North-Holland, 1980.

\bibitem{lakatos}
I.~Lakatos.
\newblock {\em Proofs and Refutations: The Logic of Mathematical Discovery}.
\newblock Cambridge University Press, 1976.

\bibitem{leroy-compiler}
X.~Leroy.
\newblock Formal certification of a compiler back-end or: programming a
  compiler with a proof assistant.
\newblock In {\em POPL '06: Conference record of the 33rd ACM SIGPLAN-SIGACT
  symposium on Principles of programming languages}, pages 42--54, New York,
  NY, USA, 2006. ACM Press.

\bibitem{lowe-new}
G.~Lowe.
\newblock Some new attacks upon security protocols.
\newblock In {\em 9th Computer Security Foundations Workshop}, pages 162--169.
  IEEE Computer Society Press, 1996.

\bibitem{maclane}
S.~Mac~Lane.
\newblock {\em Categories for the Working Mathematician}.
\newblock Springer, 1971.

\bibitem{martin-lof-itt-predicative}
P.~Martin-L{\"o}f.
\newblock An intuitionistic theory of types: Predicative part.
\newblock In H.~Rose and J.~Shepherdson, editors, {\em Logic Colloquium '73},
  Studies in Logic and the Foundations of Mathematics 80, pages 73--118.
  North-Holland, 1975.

\bibitem{martinlof85}
P.~Martin-Lof and Z.~A. Lozinski.
\newblock Constructive mathematics and computer programming.
\newblock {\em Philosophical Transactions of the Royal Society of London.
  Series A, Mathematical and Physical Sciences}, 312(1522):501--518, 1984.

\bibitem{Mathias2002}
A.~R.~D. Mathias.
\newblock A term of length 4,523,659,424,929.
\newblock {\em Synthese}, 133(1):75--86, Oct 2002.

\bibitem{mccune-solution}
W.~McCune.
\newblock Solution of the {Robbins} problem.
\newblock {\em Journal of Automated Reasoning}, 19(3):263--276, 1997.

\bibitem{nicely-pentium-fdiv}
T.~R. Nicely.
\newblock Pentium {FDIV} flaw, 2011.
\newblock FAQ page online at
  \url{http://www.trnicely.net/pentbug/pentbug.html}.

\bibitem{nipkow-flyspeck-tame}
T.~Nipkow, G.~Bauer, and P.~Schultz.
\newblock Flyspeck {I:} tame graphs.
\newblock In U.~Furbach and N.~Shankar, editors, {\em Automated Reasoning,
  Third International Joint Conference, {IJCAR} 2006, Seattle, WA, USA, August
  17-20, 2006, Proceedings}, LNAI 4130, pages 21--35. Springer, 2006.

\bibitem{paulson-natural}
L.~C. Paulson.
\newblock Natural deduction as higher-order resolution.
\newblock {\em Journal of Logic Programming}, 3:237--258, 1986.

\bibitem{paulson-found}
L.~C. Paulson.
\newblock The foundation of a generic theorem prover.
\newblock {\em Journal of Automated Reasoning}, 5(3):363--397, 1989.

\bibitem{paulson700}
L.~C. Paulson.
\newblock {Isabelle}: The next 700 theorem provers.
\newblock In P.~Odifreddi, editor, {\em Logic and Computer Science}, pages
  361--386. Academic Press, 1990.

\bibitem{paulson-blast}
L.~C. Paulson.
\newblock A generic tableau prover and its integration with {Isabelle}.
\newblock {\em Journal of Universal Computer Science}, 5(3):73--87, 1999.

\bibitem{paulson-tls}
L.~C. Paulson.
\newblock Inductive analysis of the {Internet} protocol {TLS}.
\newblock {\em ACM Transactions on Information and System Security},
  2(3):332--351, Aug. 1999.

\bibitem{paulson-consistency}
L.~C. Paulson.
\newblock The relative consistency of the axiom of choice --- mechanized using
  {Isabelle/ZF}.
\newblock {\em LMS Journal of Computation and Mathematics}, 6:198--248, 2003.
\newblock \url{http://www.lms.ac.uk/jcm/6/lms2003-001/}.

\bibitem{paulson-incompl-logic}
L.~C. Paulson.
\newblock A machine-assisted proof of {G\"odel's} incompleteness theorems for
  the theory of hereditarily finite sets.
\newblock {\em Review of Symbolic Logic}, 7(3):484--498, Sept. 2014.

\bibitem{paulson-three-years}
L.~C. Paulson and J.~C. Blanchette.
\newblock Three years of experience with sledgehammer, a practical link between
  automatic and interactive theorem provers.
\newblock In G.~Sutcliffe, S.~Schulz, and E.~Ternovska, editors, {\em 8th
  International Workshop on the Implementation of Logics, {IWIL} 2010,
  Yogyakarta, Indonesia, October 9, 2011}, volume~2 of {\em EPiC Series in
  Computing}, pages 1--11. EasyChair, 2010.

\bibitem{paulson-gr}
L.~C. Paulson and K.~Grabczewski.
\newblock Mechanizing set theory: Cardinal arithmetic and the axiom of choice.
\newblock {\em Journal of Automated Reasoning}, 17(3):291--323, Dec. 1996.

\bibitem{robinson65}
J.~A. Robinson.
\newblock A machine-oriented logic based on the resolution principle.
\newblock {\em J.~ACM}, 12:23--41, 1965.

\bibitem{rota-indiscrete-thoughts}
G.-C. Rota.
\newblock {\em Indiscrete Thoughts}.
\newblock Springer, 2009.

\bibitem{rudnicki-algebra}
P.~Rudnicki, C.~Schwarzweller, and A.~Trybulec.
\newblock Commutative algebra in the {Mizar} system.
\newblock {\em Journal of Symbolic Computation}, 32(1):143--169, 2001.

\bibitem{turing1936a}
A.~M. Turing.
\newblock On computable numbers, with an application to the
  {E}ntscheidungsproblem.
\newblock {\em Proceedings of the London Mathematical Society}, 2(42):230--265,
  1936.

\bibitem{hottbook}
T.~{Univalent Foundations Program}.
\newblock {\em Homotopy Type Theory: Univalent Foundations of Mathematics}.
\newblock \url{https://homotopytypetheory.org/book}, Institute for Advanced
  Study, 2013.

\bibitem{voevodsky-origins}
V.~Voevodsky.
\newblock The origins and motivations of univalent foundations.
\newblock {\em The Institute Letter}, pages 8--9, Summer 2014.
\newblock Online at \url{https://www.ias.edu/ideas/2014/voevodsky-origins}.

\bibitem{wadler-propositions}
P.~Wadler.
\newblock Propositions as types.
\newblock {\em Commun. ACM}, 58(12):75--84, Nov. 2015.

\bibitem{wenzel-type}
M.~Wenzel.
\newblock Type classes and overloading in higher-order logic.
\newblock In E.~L. Gunter and A.~Felty, editors, {\em Theorem Proving in Higher
  Order Logics: {TPHOLs} '97}, LNCS 1275, pages 307--322. Springer, 1997.

\bibitem{wenzel-isabelle/isar}
M.~Wenzel.
\newblock {Isabelle/Isar} --- a generic framework for human-readable proof
  documents.
\newblock {\em Studies in Logic, Grammar, and Rhetoric}, 10(23):277--297, 2007.
\newblock From Insight to Proof --- Festschrift in Honour of Andrzej Trybulec.

\bibitem{principia}
A.~N. Whitehead and B.~Russell.
\newblock {\em Principia Mathematica}.
\newblock Cambridge University Press, 1962.
\newblock Paperback edition to *56, abridged from the 2nd edition (1927).

\bibitem{woollaston-wannacry}
V.~Woollaston.
\newblock {WannaCry} ransomware: What is it and how to protect yourself, May
  2017.
\newblock Online at
  \url{http://www.wired.co.uk/article/wannacry-ransomware-virus-patch}.

\bibitem{zhan-fundamental}
B.~Zhan.
\newblock {\em Formalization of the Fundamental Group in Untyped Set Theory
  Using Auto2}, pages 514--530.
\newblock In Ayala-Rinc{\'o}n and Mu{\~{n}}oz \cite{itp-2017}, 2017.

\end{thebibliography}

\end{document}